# Machine Learning-Assisted Intrusion Detection for Enhancing Internet of Things Security

Mona Esmaeili[1], Morteza Rahimi[2,*], Hadise Pishdast[3], Dorsa Farahmandazad[4], Matin Khajavi[5], Hadi Jabbari Saray[6]

1.   Department of Electrical & Computer Engineering, University of New Mexico, Albuquerque, NM, USA
2.   School of Computing and Information Sciences, Florida International University, Miami, FL, USA
3.   College of Engineering and Computer Science, Florida Atlantic University, Boca Raton, FL, USA
4.   College of Business, Florida Atlantic University, Boca Raton, FL, USA
5.   Foster School of Business, University of Washington, Seattle, WA, USA
6.   Department of Computer Engineering, Islamic Azad University, Urmia Branch, Urmia, Iran

## Abstract

Attacks against the Internet of Things (IoT) are rising as devices, applications, and interactions become more networked and integrated. The increase in cyber-attacks that target IoT networks poses a considerable vulnerability and threat to the privacy, security, functionality, and availability of critical systems, which leads to operational disruptions, financial losses, identity thefts, and data breaches. To efficiently secure IoT devices, real-time detection of intrusion systems is critical, especially those using machine learning to identify threats and mitigate risks and vulnerabilities. This paper investigates the latest research on machine learning-based intrusion detection strategies for IoT security, concentrating on real-time responsiveness, detection accuracy, and algorithm efficiency. Key studies were reviewed from all well-known academic databases, and a taxonomy was provided for the existing approaches. This review also highlights existing research gaps and outlines the limitations of current IoT security frameworks to offer practical insights for future research directions and developments.

**Keywords:** Internet of Things, Machine Learning, Intrusion Detection, Security, Review.

---

*Corresponding author:* Morteza Rahimi (Email: mrahi011@fiu.edu)

# 1-Introduction

The IoT market has grown by 200 percent from 2006 to 2021 because of "coupling IT devices on the Internet that can send and receive data regularly" [2]. Materials and products connected to the Internet of Things have become more complex. Internet of Things technology is currently being applied to various industries, including residential, education, banking, infrastructure, smart cities, tourism, and even transportation [93, 94]. Institutions, analysts, and individuals need to take into account the unique security and availability of IoT equipment and networks to address the accelerated marketing phenomenon. The omission could negatively affect users of the IoT and disrupt the complex ecosystem. Fraudsters can access intelligent homes remotely, and intelligent automobiles can be used to terrorize residents from afar [3]. Through wired (e.g., Ethernet, WiFi) and wireless (e.g., RFID, Zigbee, WiFi, Bluetooth, 3G/4G) technologies, low-power, low-processing-power sensors can interact with other devices (e.g., gateways, cloud servers) [4, 5].

Several IoT devices are used for wearables (fitness trackers, smartwatches), home automation (lighting systems, thermostats, cameras, and locks), as well as industrial automation (process control, safety monitoring, and equipment management). According to predictions, by 2025, there will be nearly 41 billion Internet-connected devices around the world [6, 7]. Malicious software is increasingly targeting devices for several reasons, including a lack of security updates for legacy devices connected to the Internet, a lack of security priorities during development, and weak login credentials. To illustrate how an injection attack affects IoT applications, it is necessary to take a closer look at the framework of the IoT paradigm. Figure 1 shows the IoT architecture. The first layer has four basic components: perception, network, middleware, and applications. Perception is responsible for integrating intelligent devices with the framework. Cloud nodes and IoT devices communicate through the network layer. IoT data is abstracted from user applications through the middleware layer, and analytics capabilities, device management, and user reporting are supported [8-10].

Two types of intrusion detection systems (IDS) exist: based on signatures and based on anomalies. IDSs that use signatures detect attacks based on previously known attacks. These IDSs do not detect novel attack types and zero-day attacks. Anomaly-based IDSes, on the other hand, gather information on lawful user activity to determine if those users are legitimate or malicious; consequently, these IDSes can detect undiscovered attacks [11]. While anomaly-based IDSes are often more likely to generate false positives than false negatives, signature-based IDSes are more likely to create false negatives [12]. Anomaly-based techniques require collecting and monitoring data regarding the system's regular operation in the training phase to develop a model of legitimate users' behavior. To determine if a behavior is authentic or abnormal, we compare it to the model. The exponential increase in cyber-attacks has increased the need for enhanced Intrusion Detection Systems (IDS).

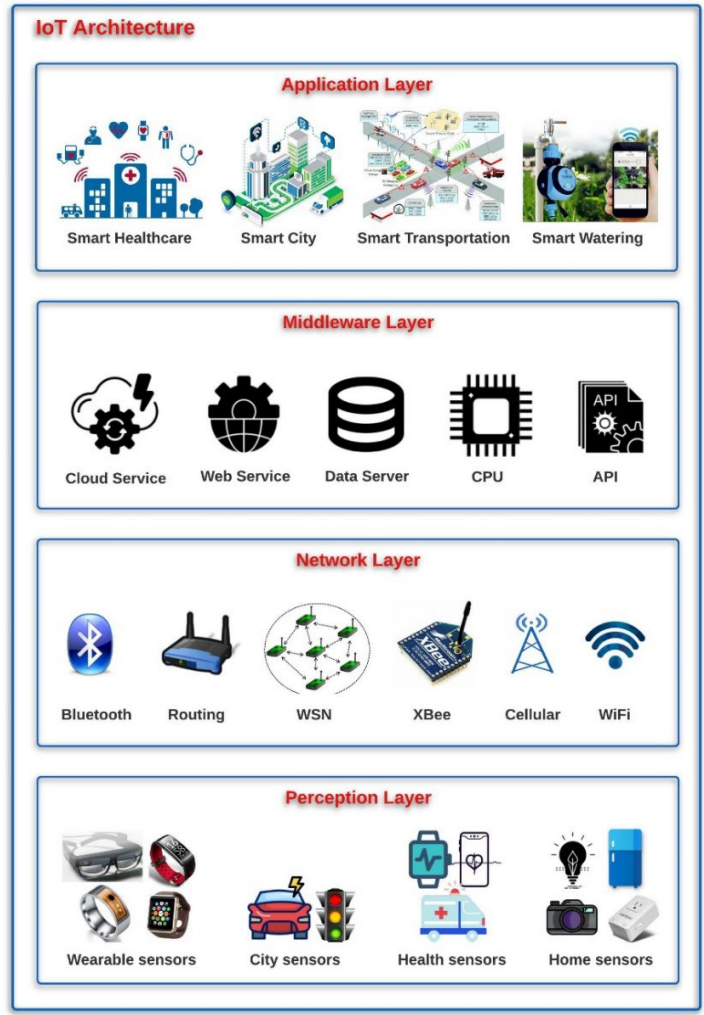

**Fig. 1.** Intrusion Detection Systems [11]

Methods such as Machine Learning (ML) are crucial since they allow for early detection of intrusions inside the system [5, 7]. Despite the plethora of alternatives, selecting the best approach is difficult due to many algorithms and optimization methods in IoT [13]. Deep Learning (DL) encompasses the three terms Machine Learning (ML), Artificial Intelligence (AI), and Artificial Intelligence (AI) simultaneously. Figure 2 shows them as subfields of each other.

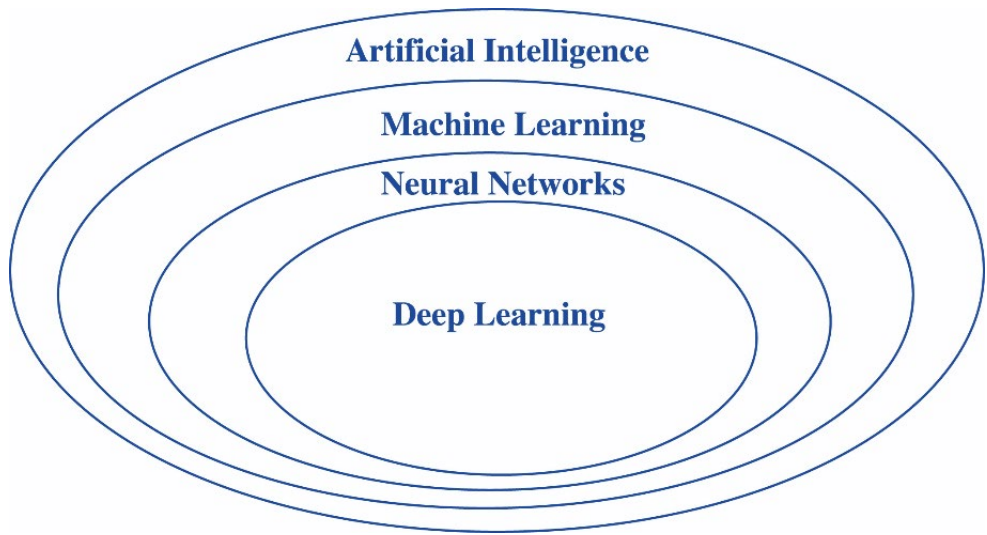

**Fig. 2.** Relationship between AI, ML, ANN, and DL.

There are specific connections between ANN, deep learning, and machine learning. Using computers to study human cognition is analogous to studying biological processes [14, 15]. In machine learning, algorithms are used to assess data relevant to a specific issue (e.g., intrusion detection), learn from it, and then use that knowledge to identify patterns that help solve the issue. Neurons are the building blocks of neural networks (NN). A neural network simulates the behavior of a real neuron by sending data across links of neurons. Using a weighted average of their inputs, neural neurons affect the behavior of real neurons. Theoretically, researchers [16] describes NN as a "massively parallel aggregation of basic processing units that learn from their surroundings and store that information in their connections." Machine Learning (ML) has proven helpful in addressing a variety of risks in the Internet of Things (IoT) environment, mainly when applied to various IoT security aspects [17-20].

Figure 3 shows the components of an IoT system and several ways to create an intelligent IoT system. The purpose of this study is to compare machine learning-based intrusion detection systems against techniques used in IoT attacks. This essay is structured as follows. Section 2 reviews various publications on this topic from 2015 to 2022. In Section 3, we describe intrusion detection and anomaly detection systems for IoT, identify IoT botnet attacks, and use machine learning to detect IoT attacks. We conclude our study by discussing its limitations and making several recommendations for future study.

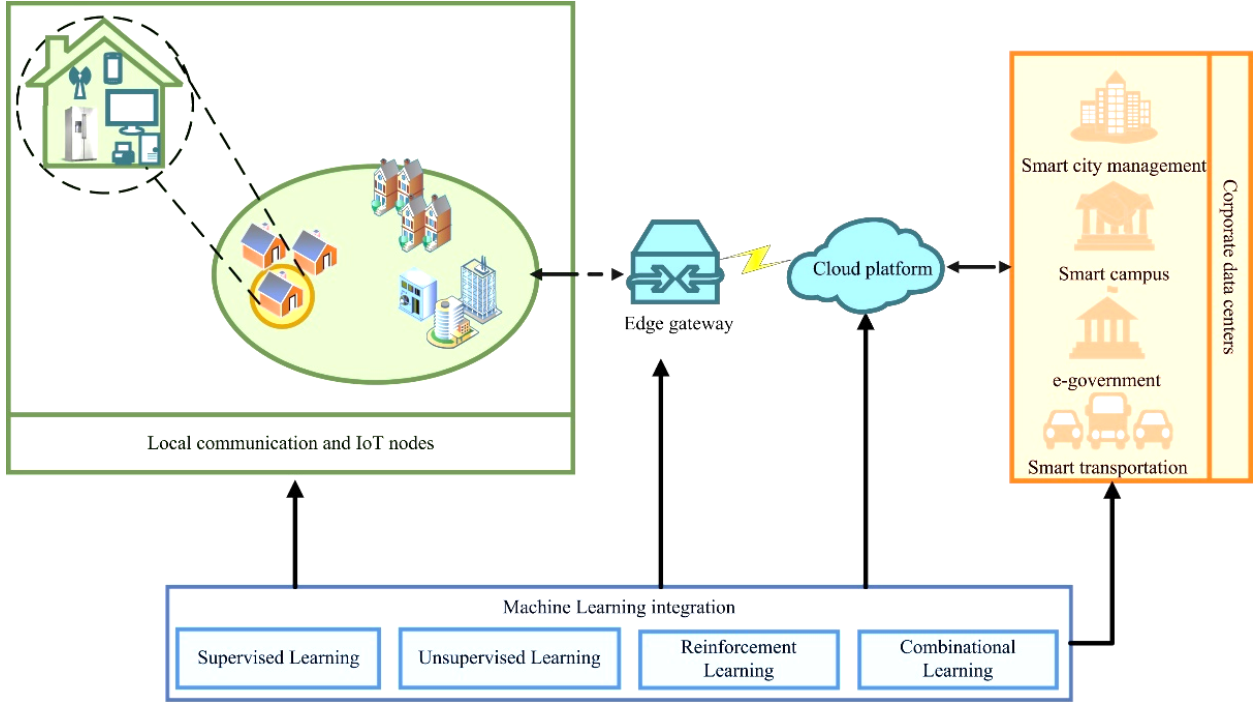

**Fig. 3.** Intelligent IoT systems at different stages of development: machine learning approach [21]

## 2- Literature review

This section discusses how to protect settings powered by machine learning and the Internet of Things by utilizing intrusion detection and identification technologies provided by various authors. Literature reviews were conducted on the topic. The IEEE and ACM Digital Libraries, as well as Elsevier and Springer databases, were searched. Here, we discuss and compare the articles found during the review. Table 1 is an overview of intrusion detection systems for the Internet of Things [21]. It is possible to determine which techniques are capable of multiclass detection and their primary shortcomings. A detection method developed by Hagos et al. [22] Support Vector Machines (SVMs) and the Least Absolute Shrinkage and Selection Operator are combined in this algorithm (LASSO). As part of an IoT context, Diro and Chilamkurti [23] propose using Deep Neural Networks (DNN) as a distributed solution for detecting intrusions. Detection methods are dispersed throughout the fog layer. A DNN detection model is trained across each distributed node, resulting in a model for each node. To verify the technique, the NSL-KDD dataset was used. In [24], the authors present a hybrid technique for binary intrusion detection in fog computing settings using Deep Neural Networks and K-nearest neighbors (KNN). Additionally, we describe the framework that underpins this work. Vinayakumar et al. [25]

also examined deep networks for intrusion detection. They presented their method using Deep Belief Networks (DBN).

### 2-1-Machine learning-based work

Kolias et al. [26] employed a variety of well-known machine learning algorithms (AdaBoost, OneR, J48, Naive Bayes, Random Forest, ZeroR, and Random Tree) to find the best classifier for recognizing intrusions in the AWID dataset. As a result, J48 achieved 96.20 percent accuracy when all 154 characteristics were used and 96.26 percent accuracy when only 20 characteristics were used. Moreover, Thanthrige et al. [27] suggest selecting features using information gain and Chi-Squared statistics. A variety of feature vectors were then employed (111, 41, and 10). The performance test showed that using 41 features, Random Tree was able to raise accuracy from 92.17 percent to 95.12 percent.

Aminanto et al. [28] presented a method for detecting impersonation attacks in wireless networks. An ANN, a Decision Tree, and an SVM algorithm were applied to the feature selection process. Using the SVM approach to train 11 features, the best performance was 99.86 percent. The team did not evaluate other attacks (injections and floodings). Kaleem et al. [29] proposed a method for ranking cognitive features and an ANN classifier for IDS. The method categorizes incidents based on whether they are considered assaults or normal incidents. They use a cognitive feature ranking approach that utilizes feature space analysis. They pruned the input neurons to remove unimportant neurons in the ANN's first layer. This strategy improved accuracy from 97.84 percent to 99.3 percent when the number of characteristics was reduced from 154 to 6.

In response to the withholding attack, Yujin et al. [30] proposed the fork assault. The fork assault is always more successful than the withholding attack. 51 percent of attacks occur when an attacker controls more than 50% of a network's hashing power. An attacker who controls more than 50% of the computing power on a blockchain network can prevent transactions from being confirmed. As a defense against this attack, Bastiaan et al. [31] investigated Two-Phase Proof-of-Work (2P-PoW). Double-spending assaults occur when a peer node creates two transactions and sends them to two receivers simultaneously. One-time signatures are used to solve this problem [32].

Distributed denial-of-service (DDoS) assaults are among the most common attacks against blockchain networks. According to various research, attackers often target the information flow between two peers in a blockchain network [33]. These studies did not provide any mitigating techniques. Apart from producing cheap transactions in the mempool, another way for conducting a distributed denial-of-service attack against a blockchain network is possible [34]. This is referred to as a penny-flooding assault. Kumar et al. [35] investigate blockchain stress by assessing the vulnerabilities revealed by distributed ledger technologies.

Kumar et al. [36] created an IDS for a blockchain-based Internet of Things (IoT) system. The authors proposed a federated learning-based IoT-based attack detection system. They accomplished this by using the PySyft package. However, crucial evaluation criteria, such as the false alarm rate, were missing from this research. Bakhsh et al. [37] suggested an adaptive IDS for IoT devices that makes use of agent technology to provide portability, rigidity, and self-starting behaviors. This hybrid system identified both abuse and irregularities by using both host-based and network-based capabilities. This research lacked performance data for the IDS or its operating systems. Anthi et al. [38] suggested an adaptive and predictive IDS system for IoT contexts based on signatures and anomaly detection. Due to the study's imprecise model assumptions, it was only possible to identify a small amount of assault. We determine the predicted threat model and attack aim in a blockchain-enabled IoT network from the literature (see Tables 1 and 2).

**Table 1:** An overview of intrusion detection systems for IoT

| Proposals | Implementation Strategy | Method | Threats to security | Solution for Validation |
|---|---|---|---|---|
| [39] | Centralized | Anomaly-based | Man-in-the-middle | Modeling |
| [40] | - | According to the sign | - | - |
| [41] | multiple | Consistent with specs | Attacks on multiple targets | - |
| [42] | - | Consistent with specs | DoS | Modeling |
| [43] | Centralized | According to the sign | DoS | Research-based |
| [44] | Centralized | - | Attacks on multiple targets | Modeling |
| [45] | multiple | hybrid | Attacks on multiple targets | Modeling |
| [46] | - | Anomaly-based | - | - |
| [47] | Centralized | According to the sign | - | Hypothetical |
| [48] | multiple | Consistent with specs | - | Research-based |
| [49] | Distributed | According to the sign | Common attacks (Snort and Calmv databases) | Research-based |
| [50] | Distributed | Anomaly-based | DoS | Modeling |
| [51] | - | hybrid | Attacks on multiple targets | Modeling |
| [52] | Distributed | hybrid | Attacks on multiple targets | Modeling |
| [53] | - | Anomaly-based | traditional | Research-based |
| [54] | multiple | Anomaly-based | - | - |
| [55] | multiple | Anomaly-based | Attacks on multiple targets | simulation |

**Table 2:** Machine learning background

| Author | An analysis of the gaps |
|---|---|
| *IoT methodologies* | |
| Zarpelão et al. [56] | The article discovered a power usage difference between machine learning and IDS. |
| Rehman et al. [57] | The article highlighted a need for RAOF that ML and IDS can fill based on power usage and hop count. |
| Le et al. [58] | Combinations of attacks have not been considered. A novel approach to anomaly-based detection might be based on power consumption and dropped packets. |
| *MRHOF and OFO attacks* | |
| Airehrour et al. [59] | Neither OFO nor MRHOF was found to be under attack. Each IoT combination attack will be based on MRHOF and OF0. |
| Airehrour et al. [60] | MRHOF and OF0 cannot detect and isolate Rank and Sybil attacks. |
| Mehta and Parma [61] | According to the paper, there is a need to detect possible OF attacks. |
| *Methodologies and features used in IDS* | |
| Sheikhan and Bostani [62] | Based on misuse-based detection, the research failed to detect unknown attacks. |
| Mayzaud et al. [63, 64] | Authors claim that their research is a feasible way to detect anomalous behavior in IoT devices, but there is no evidence that additional attacks beyond DAG can be detected. |
| Lee et al. [65] | When describing detecting malicious behavior using power consumption and network traffic, the terms OF and MRHOF are omitted. |
| Sousa et al. [66] | Simulated OF-FL, CAOF, and other significant OF routing metrics were omitted. |
| Napiah et al. [67] | To make ML algorithms more efficient, features were reduced from 77 to 5, removing power consumption. |
| *Datasets and ML classifiers* | |
| Haq [68] | ML-IDS development should take into account 49 studies reviewed in the paper. Methods and algorithms for machine learning, datasets, and feature selection. |
| Nannan et al. [69] | A high false alarm rate was identified in research on anomaly detection. Methods for applying ML, algorithms for classifiers, datasets, and feature selection |
| Buczak and Guven[70] | Since KDD 1999 was produced, several attacks have taken place. These include Internet of Things attacks. |
| Alam et al. [71] | According to the paper, there has been little research on applying conventional ML methods to IoT datasets. Identify or develop a unique dataset for IoT features and attacks, including the ML approach, algorithm, dataset, and feature selection |
| *Techniques for preprocessing and balancing* | |
| Yin and Gai [72] | An imbalanced dataset should be developed considering 12 datasets. |

## 3- Intrusion Detection in the Internet of Things

Internet of Things solutions are insufficient because of the unique systems associated with the Internet of Things, which affect the development of intrusion detection systems. In the beginning, you must analyze the network nodes' memory and processing capabilities. Most nodes on the Internet of Things have limited processing power. The Internet of Things makes it more challenging to identify nodes that can host intrusion agents. Second, network architecture should consider functional factors. The transfer of packets in a normal network is controlled by the switches and routers attached to the end systems. However, traditional nodes often serve as packet transporters and terminals in IoT networks since they are designed in multiple steps. Network protocols are also part of IoT networks. In the Internet of Things networks, protocols such as RPL LoWPAN IEEE802.15.4 and CoAP are used that are not found in conventional networks [62]. On the other hand, these studies focus on the development of intrusion detection systems for Internet of Things components. However, none of them examine key methods for penetration into the Internet of Things systems. The following sections describe how to implement intrusion detection systems and common security issues relating to the Internet of Things. We will also discuss the validation process for intrusion detection systems for the Internet of Things. According to Figure 4 [73], comparable works are grouped as follows: 1. Strategic positioning 2. Design 3. Methods of detection 4. Threats to security 5. Methods of validation.

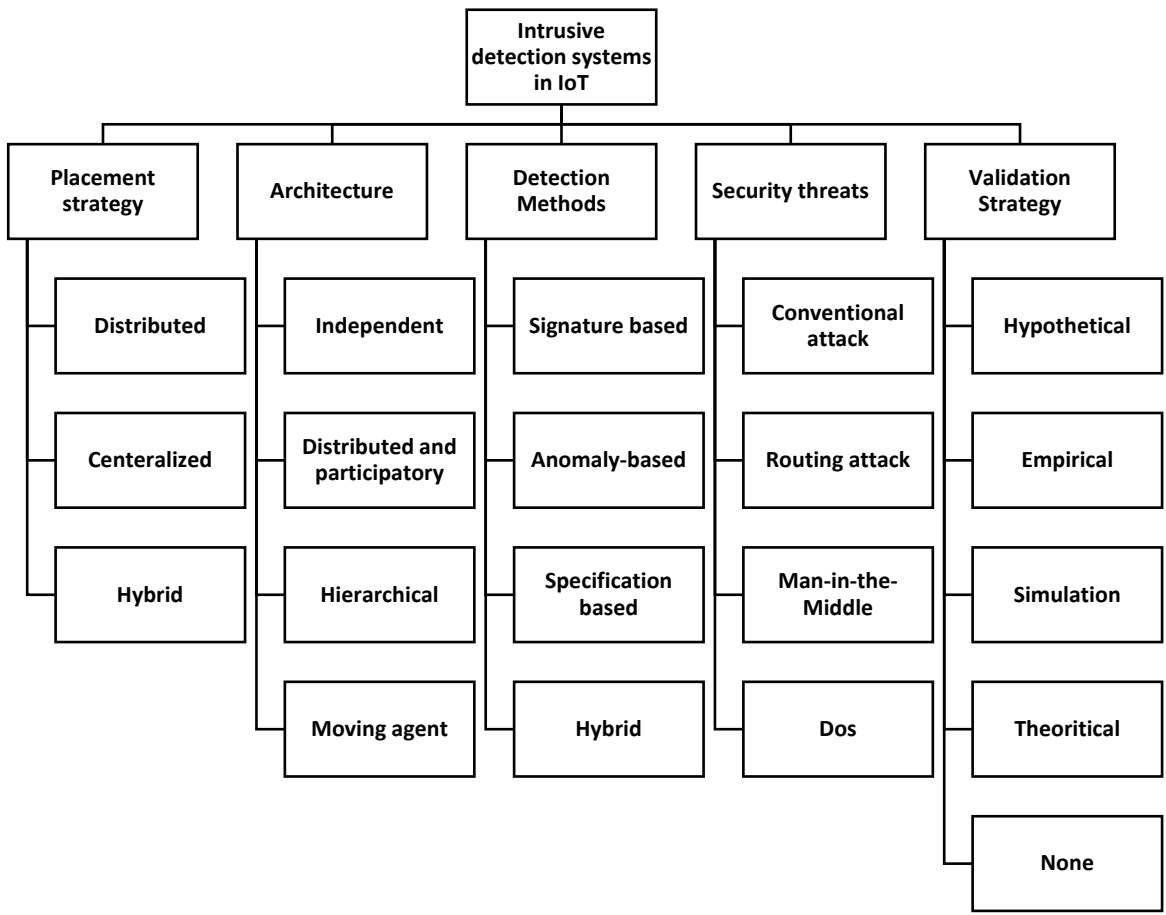

**Fig. 4.** Methods of detecting IoT intrusions

## 3.1. Large-scale attacks

The Internet of Things has increased visibility, control, and potential practically everywhere. There are many applications for IoT today, among them autonomous cars, traffic congestion, patient monitoring, high-quality health care, and smart home appliances [92]. Individuals have increased their productivity and improved their quality of life through these solutions [70-72]. Despite this, these ubiquitous networked gadgets are vulnerable to cyber-attacks due to their widespread connectivity and processing capabilities. These devices may become bots or zombies because there are no security precautions. These devices form a botnet when hundreds of millions of them become infected. Command and control (C&C) servers manage these botnets, and they are used to carry out a variety of large-scale malicious attacks. These attacks take many forms [73].

In some cases, such as TCP timeout retransmission and maintaining an open HTTP connection, exhaust the server's resources, rendering the server inaccessible to valid requests. Alternatively, some large-scale assaults rely on volume. As a result of transmitting large amounts of bandwidth, these attacks prevent legitimate queries from accessing crucial services. Attacks include DDoS, Simple Network Management Protocol (SNMP), TCP SYN, UDP flooding, and ICMP flooding [73]. In addition, it featured a category of application-layer attacks, which used fewer resources to disrupt critical systems. A few examples of such attacks include zero-day vulnerabilities, Slowloris, web server exploits, and OpenBSD attacks. By flooding the internet with traffic, DDoS attacks attempt to interrupt normal operations and deplete the server's resources or bandwidth, thereby shutting down the website. A network attack, a protocol attack, and an application attack [5] are three types of attacks. Regardless of their type, their primary objective is to interfere with harmless communications in order to prevent real users from accessing the service.

## 3.2. Techniques for detecting intrusions

The computing and storage capabilities of IoT devices are low. Because of the Internet of Things (IoT) setting's heterogeneous nature, relatively large typical behavior, and increasing vulnerability as IoT devices multiply rapidly, traditional intrusion detection systems (IDS) are insufficient [14]. According to [66], protecting these devices requires a paradigm shift. In IoT systems, two of the most commonly used strategies are signature-based (also called misuse- or knowledge-based) and anomaly-based (also called behavior-based) approaches [1][50]. Hybrid detection systems can be made by combining them, but this is time-consuming [24]. Signature-based algorithms identify traffic as benign or malicious based on preexisting threat information, while anomaly-based systems detect attacks based on traffic patterns [33]. These methods provide reasonable protection against established threats. It takes time to maintain a signature database, which is one of the drawbacks of a signature-based strategy. As a database grows, it becomes increasingly computationally expensive to compare the input to it. This technique relies on previously known attack signatures; it cannot detect zero-day attacks or new attacks. Anomaly-based detection is preferred by analyzing regular traffic patterns and alerting or restricting traffic when an aberrant way is detected. Using anomaly-based systems can effectively identify zero-days and unknown threats; however, many false positives may result [43]. An anomaly-based IDS is a key focus of research because of its robust detection capabilities for large-scale, zero-day, and unknown threats.

### 3.3. Taxonomy of IoT threats

Many threats are posed to the Internet of Things ecosystem. However, they are susceptible to large-scale and distant attacks due to their heterogeneous nature, considerable autonomy, ability to communicate and control the internet, and resource constraints. Active attacks disrupt network traffic in real-time or reduce the availability of system resources. As opposed to forceful attacks, passive attacks target IoT resources through watching, listening, eavesdropping, and analyzing traffic patterns. An Internet of Things threat taxonomy at different tiers is illustrated in Figure 5. Researchers will be able to develop tools capable of identifying various IoT attacks, including zero-day attacks. This section outlines some of the most well-known and popular threats to IoT systems that we will focus on in our study.

### 4 Machine Learning algorithms

There are two prominent types of machine learning methods widely used in the IoT environment: supervised and unsupervised. Supervised learning techniques use labeled data in training to detect abnormalities in new data samples. The training data is not labeled; rather, the learning algorithm groups/classifies it using a variety of grouping algorithms. The majority of IDSes that are signature-based use supervised learning algorithms, while those that are anomaly-based employ unsupervised learning techniques [60, 61].

Two classification labels are used in binary classification, the most basic type of classification. In binary classification, intrusions are classified as normal and assault (or abnormal, anomalous, etc.). Data can be classified using more than two labels using multiple-class classification. Data may be classified as normal, DoS, User to Root, Remote to Local, or Probing for intrusion detection purposes [65]. Learning by machine occurs when computers are presented with data and can improve their performance. Computer programmers need to automate spotting complicated patterns and generating intelligent judgments based on data. One of the most challenging aspects of machine learning is teaching the computer to automatically recognize a handwritten postal code after learning several examples from a collection of samples. In recent years, machine learning has expanded rapidly [66].

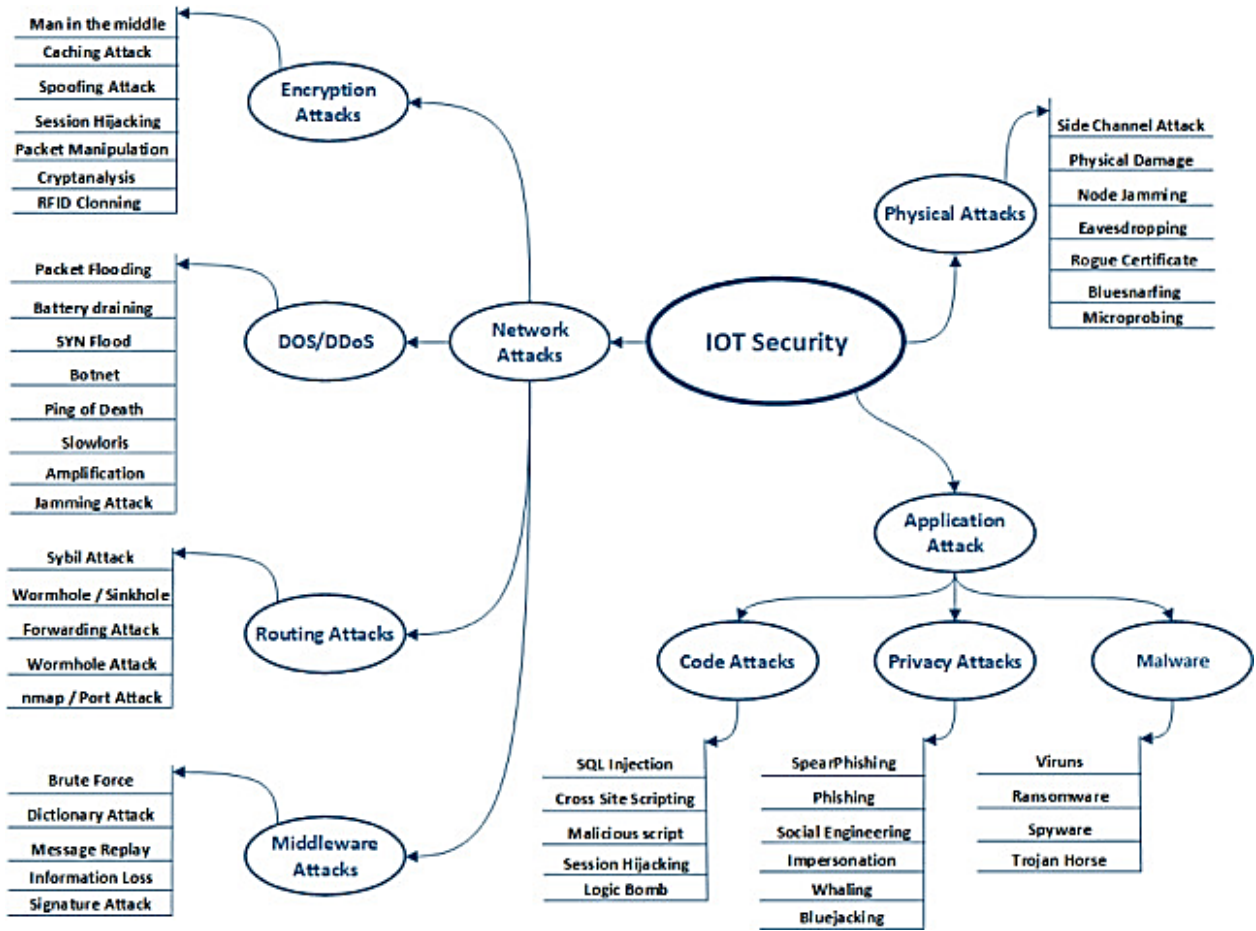

**Fig. 5:** An Internet of Things threat taxonomy at different tiers

## 4.1. Supervised learning

Essentially, supervised learning is the same thing as categorization. The training dataset contains labeled cases that teach supervision. When it comes to recognizing postal codes, for example, a set of photographs with an associated machine-readable interpretation is used as instructive examples to measure how the model is learning [50-55].

## 4.2. Clustering

Clustering is the same as unsupervised learning. Since the input examples lack class markers, the approach is called unsupervised learning. Most often, data are classified using clustering. An unsupervised learning approach, for example, accepts photographs of handwritten numbers as input in the case of postal code recognition. Suppose ten clusters are discovered in this way. Each cluster corresponds to a distinct digit between 0 and 9. The model does not convey the semantic meaning of the discovered clusters due to the unlabeled nature of the training data [18].

### 4.3. Semi-supervised learning

Semi-supervised learning employs both labeled and unlabeled data to build the model. As one example, labeled examples aid in learning class models, whereas unlabeled examples are used to correct class boundaries [19]. With two classes, a subset of cases may be categorized as positive, while the remaining are labeled as negative. Unlabeled examples may allow the decision boundary to be set more precisely. Furthermore, despite their labeling, two positive examples in the top right corner are likely skewed or noisy [19]. Computer-based active learning programs engage students in the educational process by actively encouraging them to take part. A user (for example, a subject matter expert) can tag an Instance, which may be derived from a collection of unlabeled Instances or generated by the learning program [45]. Since the model is only tested in many situations, this technique seeks to improve its accuracy by using input derived from human users. Machine learning and data mining are closely related. Similarly, machine learning depends on the correctness of models regarding classification and clustering. The importance of scalability and reliability of mining techniques for large datasets and the creation of novel and alternative methods is emphasized in data mining [11] (see Table 3).

**Table 3.** A recent survey and comparison of machine learning techniques for IoT systems.

| Ref. | Year | Optimization | KPIs | Results |
|------|------|:---:|:---:|---------|
| [74] | 2017 | ✓ | ✗ | The emphasis of the research is on self-organizing mobile networks. |
| [75] | 2018 | ✗ | ✗ | Provided solutions for the management of vehicular resources. |
| [76] | 2018 | ✗ | ✗ | Contained only solutions relating to context-aware computing. |
| [77] | 2019 | ✓ | ✓ | Concentrated exclusively on approaches based on artificial neural networks. |

| Ref. | Year | Optimization | KPIs | Results |
|---|---|---|---|---|
| [78] | 2019 | ✗ | ✗ | Security solutions for IoT systems. |
| [79] | 2018 | ✗ | ✗ | Analyzed IoT data using analytic techniques. |
| [80] | 2019 | ✓ | ✓ | Massive MTC on ultra-dense networks: methods based on focused learning. |
| [81] | 2016 | ✗ | ✗ | Detecting cyber-security intrusions using learning techniques. |
| [82] | 2017 | ✓ | ✗ | A routing solution company focused on intelligent networks. |
| [83] | 2016 | ✓ | ✗ | Study of CR networks specifically. |
| [84] | 2017 | ✗ | ✗ | Described only techniques for reducing self-interference in wireless networks. |
| [95] | 2023 | ✓ | ✓ | Investigated various optimization techniques, such as genetic algorithms and ML models, for power conservation and security enhancement in IoT-assisted smart systems. It also assessed KPIs associated with energy usage, data privacy, and security metrics, |
| [96] | 2024 | ✓ | ✗ | Explored optimization in adversarial ML attacks, especially optimization-based poisoning attacks and their defense mechanisms. |
| [97] | 2024 | ✗ | ✗ | Focused on security vulnerabilities in IoT environment. It primarily deals with open-source libraries and threat models. |
| [98] | 2022 | ✗ | ✗ | Investigated ML models for malware detection in IoT enterprise systems. Also, highlighted security issues. |
| [99] | 2023 | ✓ | ✓ | Evaluated ML-based intrusion detection systems and discussed optimization techniques for secure communication and network throughput. KPIs such as detection accuracy, false positive rates, and error rates are mentioned for evaluating intrusion detection systems. |
| [100] | 2021 | ✓ | ✗ | Investigated optimization methods such as hyperparameter tuning and real-time responsiveness in deep learning models for IoT security. |
| [101] | 2022 | ✗ | ✗ | Focused on security, privacy, and safety in IoT using ML. |

| Ref. | Year | Optimization | KPIs | Results |
|-------|------|:---:|:---:|---------|
| [102] | 2021 | ✓ | ✗ | Assessed optimization methods in neural network models for IoT security, specifically related to computational complexity and energy consumption. |
| [103] | 2023 | ✓ | ✗ | Investigated optimization in security feature selection for IoT systems. The focus was on reducing data dimensions for efficient threat detection. |

## 5. A method for developing algorithms for different key performance indicators (KPIs)

This section will study machine learning methodologies in connection with a range of key performance indicators (KPIs) for IoT applications. As seen in Figure 6, We will discuss the metrics used in the literature to assess the performance of different machine learning algorithms for IoT devices.

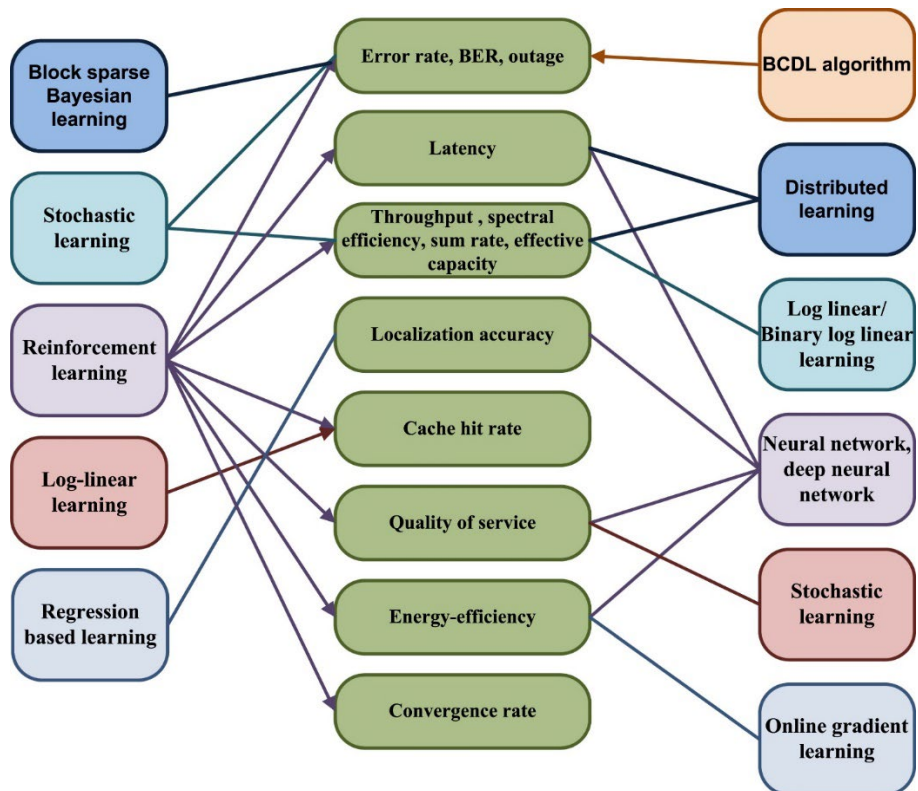

**Fig. 6**. IoT systems for different KPIs.

### 5.1. Classification

Data classes and ideas are classified using a model (or a function) that describes and separates them. The model is built based on the evaluation of experimental outcomes (data objects with class labels). Undefined data objects are annotated with class tags using the model. As part of the categorization procedure for data (during which the constructed model is used to predict the label for the given data class), the learning step (during which a classification model is constructed) and the classification stage (during which the data is classified) are also included [18]. During the first phase, a classifier is constructed to represent the sets of data or ideas to be analyzed. An algorithm is used to generate this learning step (or learning phase) based on a collection of database instances and class tags. The model is used in the second procedure to categorize the data. To apply the model to original data, the model must first be assessed for accuracy. This is done through a series of trials. Models may be developed using a variety of techniques, including categorization principles, decision trees, mathematical formulae, and neural networks [34-38].

## 5.2. Evaluation Metrics

When it comes to evaluating the performance of a classifier, machine learning provides a wide range of measures. It is crucial to choose the appropriate performance indicators based on the application's actual needs. A classifier may perform well on one step, while on another, it might not [86]. In model assessment, one of the objectives is to extract the most value from performance measurements as possible. Based on our evaluation of research articles, we isolated performance learning performance indicators.

## 5.3. Accuracy (ACC)

For classifiers, accuracy is a common performance metric. A classifier is evaluated based on its ability to recognize intrusions or attacks from an IDS perspective. Infiltration attempts are measured as the percent of total accurately identified inputs [24].

## 5.4. Precision (PR)

The accuracy of a model can be an indicator of its effectiveness, but it cannot be the only judgmental factor. Inequalities in datasets create uncertainty. While a model may show high accuracy scores when the input dataset is constant, it fails to do so when the input changes and performs poorly [16]. PR (Performance Ratio) is used instead in these situations. A true positive rate is the proportion of those anticipated as positive among those who are positive. In other words, it refers to the percentage of malicious packets correctly identified [14]. The higher the accuracy score, the better the model can categorize the attack data. According to [11], precision refers to whether or not truly favorable outcomes are certain.

## 5.5. Recall (R)

In this recall, we are looking at sensitivity and whether all positive examples were correctly forecasted. As a result, it can correctly identify a small fraction of malicious packets. If a model is incapable of detecting large-scale attacks, malicious traffic may not be detected. System security will be compromised as a result.

## 5.6. F-measure (F1)

The accuracy and performance of a model are determined by precision and recall. By reducing false positives and false negatives, a model with a high F-measure score recognizes attack traffic successfully. Accuracy and recall are symmetrized in F-measure.

## 6. Decision tree-based classification

The leaf node of a decision tree corresponds to a class label, while the inside nodes represent a decision or a chance. Branch nodes of a decision node indicate the outcome of the evaluation of each feature. The ID3 and C4.5 algorithms are well-known decision tree algorithms. In both methods, information entropy is used to construct decision trees from training datasets for labeling data. For the most part, decision trees are accurate at classifying data. Combined with decision trees, clustering techniques help reduce processing time [5].

Figure 7 shows a decision tree for identifying TCP connections. The rectangular rectangles indicate characteristics, while the branches indicate regulations. The NSL-KDD dataset describes these characteristics for TCP connections. A flag is associated with each TCP logged-in feature. The value of the flag is either zero or one. An attempt at login that ends in a zero indicates a failed attempt. An attempt at login that ends in one indicates a successful effort. The error rate measures how many TCP connections from the same host fail with a Syntax Error (SYN).

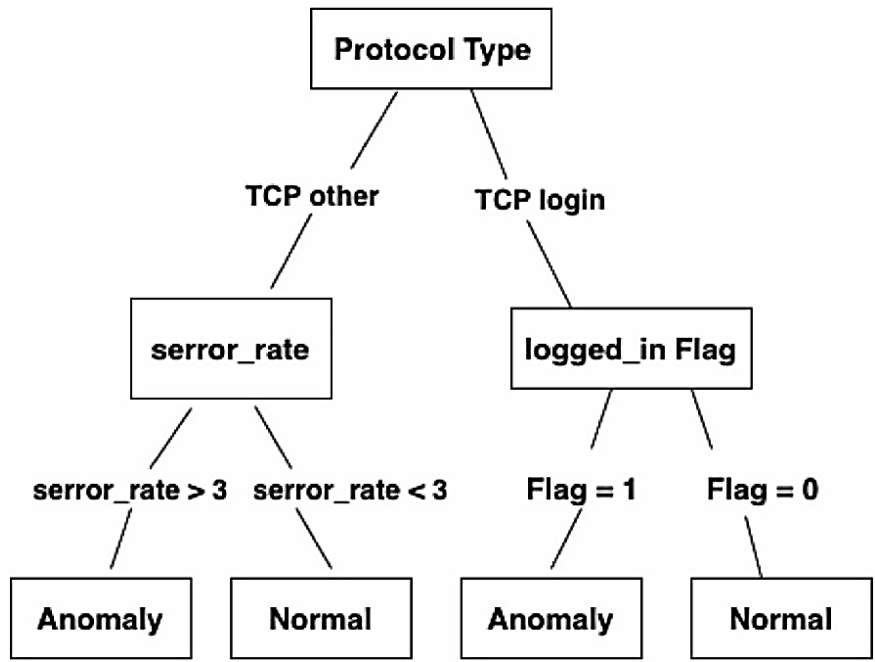

**Fig. 7.** Decision Trees (DTs)

## 7- Dataset

### 7.1. KDD99

A benchmark dataset for evaluating various IDSes is KDD (Discovering knowledge through databases), released in 1996. The dataset was compiled by Stolfo et al. [85] .According to [21] DARPA 98, the KDD99 dataset was constructed. The KDD99 dataset consists of 41 attributes classified as either normal or attack. Twenty-one different assault types are included in the dataset. In addition to Denial of Service (DoS) attacks, User to Root (U2R) attacks, Remote to Local (R2L) attacks, and Probing attacks, there are several other kinds of attacks. There are 78 percent and 75 percent duplicate entries in the training and testing datasets, respectively. A machine learning model cannot be trained using U2R (52 attacks in the training dataset) and R2L (1106 attacks in the training dataset) attacks, resulting in an inbalanced KDD99 dataset.

### 7.2. NSL-KDD

In addition to NSL-KDD, another dataset that is used a lot is Network Socket Layer-Knowledge Discovery in Databases (NSL-KDD). This dataset was derived from the KDD99 dataset by Tavallaee et al. [86]. This dataset does not contain duplicate entries, making it a better candidate for testing machine learning-based IDSes than KDD99 [21]. The distribution of data across different classes is asymmetric, though, as with KDD99.

### 7.3. UNSW-NB15

A new dataset for evaluating IDSs was suggested by Moustafa et al. [87] in 2015. As faithfully as possible, this dataset captures network activity. Tcpdump was used to capture raw network traffic. A number of critical features were derived from traffic data by Argus and Bro-IDS [28, 29]. There are 49 distinct features in the dataset and 2,540,044 unique entries. Fuzzers, Analysis, Backdoors, Denial of Service, Exploits, Generic, Reconnaissance, Shellcode, and Worms are among the nine attack types in this collection. For Analysis, Backdoor, Shellcode, and Worm, only a few samples are available for training.

## 8. Techniques for detecting intrusions based on machine learning

A critical review of papers published in the past ten years on ML techniques' intrusion detection is presented here. The bulk of these articles used supervised learning techniques for binary or multiclass categorization. They evaluated their methods using a variety of publicly available benchmark datasets. As we discussed in section 2.3, there are several widely used datasets. Accordingly, we classify papers according to how they label datasets (binary or multiclass) and the datasets used to assess the paper's approaches.

## 8.1. Intrusion detection system architectures

This section [20] describes how the detection system's design enables automatic, distributed, participatory, hierarchical, and agent-based classification. Every observer node identifies intrusions independently and gathers data independently. This could be centralized or distributed. Each network node operates as an observer node in a centralized observer node architecture. In a distributed observer node design, each sensor node must be included in at least one observer node, and each observer node monitors a different part of the network. A separate intrusion detection system is installed in each monitoring node.

- **Distributed architecture:** intrusion detection system agents are executed on each monitoring node. Monitoring nodes collaborate to detect intrusions. Data and warnings shared across the network with another monitoring node contribute to the conclusion of the intrusion detection system as it monitors its surrounding nodes. Diagnostic performance is improved with this method. An infrastructure that adheres to the DODAG standard can benefit from this design.
- **Hierarchical architecture:** compatible with hierarchical sensor networks. Multiple DODAGs (with a common sink node) are supported. The sink nodes serve as cluster head agents for an intrusion detection system. In contrast, the local agents are created and deployed according to the system's autonomous architecture and collaborate in the intrusion detection process.
- **Moving agent architecture:** Collaborates on detecting intrusions between various mobile agents. Agent mobility may enhance the performance of intrusion detection systems. Mobile agents are executable pieces of code that travel from one node to another as a part of a self-controlled program. While an agent migrates from one node to another, computations are also performed.
- **Examines several routing attacks:** It is including sinkholes, selective send, hello floods, wormholes, identifier attacks, and Sybil. Walgren et al. [88] have developed an intrusion detection system confined to selective sending of attacks. Reza et al. [89] have developed a technique to detect sinkholes and selective sending attacks. Cervantes et al. [90] developed a method for identifying sinkhole attacks. The authors of this study focus on node mobility and self-repair, which are closely related to [91].

## 8.2. Limitations of Deep Learning (DL)

The use of deep learning to identify large-scale threats in IoT requires identifying hidden patterns in data through a multi-layered approach. DL makes it possible to use state-of-the-art methodologies, increase efficiency, and achieve previously unattainable results due to classical machine learning algorithms. Researchers employed individual deep-learning methods such as CNNs, ANN, MPLs, LSTMs, RNNs, and CNNs. When applied to various deep learning algorithms, these strategies do have drawbacks; despite providing useful performance metrics in general, deep learning approaches have three significant drawbacks: (1) highly linked layers lack intra-layer connections; (2) DL layered networks must be manually tuned to get optimum results; and (3) big DL models use a significant amount of energy and computational resources [19]. In addition, deep learning models are subject to training time constraints as the amount of training data grows [12].

The use of CNNs in image and language processing is widespread [12]. Also, CNN has demonstrated impressive performance in various other applications, including detecting benign and malicious network traffic in an IoT setting [11, 12, 14, 79]. CNN-based models are computationally complex, making them difficult to implement on IoT devices with limited resources [3]. In a CNN, the complexity increases as the layer count increases and the levels are connected in large dimensions. The CNN begins to make judgments about the characteristics [63]. Natural language handling and text processing are promising applications for RNNs [77]. Because they can process sequential data, RNNs are well-suited for analyzing IoT sensor data and detecting attacks. Gradients disappearing or ballooning is one of the shortcomings of RNNs. They cannot capture long-term relationships. As a result of the timestamp's computation reliance on the previous timestamp, parallelism's capabilities are limited. As an alternative, the LSTM emerged from RNN and can detect regular and malicious traffic through pattern learning based on ling sequences [91]. Due to its ability to learn long-term dependencies and address the vanishing gradient issue, LSTM is used more extensively than RNN. In comparison to CNN and RNN, LSTMs are superior IDSs since they can learn from temporal sequences and long-term dependencies. As a result, LSTM requires a large amount of training time since it cannot calculate simultaneously at multiple timestamps.

Using deep learning models to detect threats in IoT systems, researchers have combined deep learning models to improve performance [19, 10]. An algorithm combining CNN with LSTM is presented [91] for detecting large-scale attacks. Their technique begins with a CNN layer, then moves to LSTM layers, drop-out layers, and finally, a fully connected layer. A comparison was made between the performance of hybrid models and the performance of individual models (MLPs, CNNs, and LSTMs) and classical machine learning models (SVMs, Random Forests, and Naive Bayesian models). In terms of performance, CNN and LSTM outperformed separate models. Using their suggested model, they generated accuracy, precision, and recall values that exceeded 97.16 percent. When CNN and LSTM were run independently, their accuracy was 95.14 percent and 86.34 percent.

## 8.3. Security features for IoT

Recently, academics have been focused on developing scalable, lightweight intrusion detection systems capable of detecting aberrant IoT network traffic. Despite years of study and better technologies, researchers still have difficulty managing large amounts of data and modifying network traffic patterns [62]. Feature selection plays an important role in machine learning, which impacts the effectiveness of ML IDS. To develop a successful IDS based on ML, researchers aim for a reduced feature set that doesn't impair attack classification accuracy. To train the classifier, a number of machine learning algorithms examine input data for features that represent the data set. Several examples include support vector machines (SVMs), random forests, and decision trees. Errors in network traffic patterns are evaluated by using a learned classifier [63]. An important aspect of feature selection is the input data and classifier being used. When there are many characteristics in the input data, feature selection is complicated by the lengthy computation time [64].

There are no predefined criteria for the type and quantity of classification characteristics that should be used. Researchers use the dataset to examine stateless and stateful traffic from different perspectives; some examine network traffic for abnormalities from stateful traffic, while others investigate stateless traffic. In a similar line, other studies propose manual feature selection based on personal experience. In contrast, others advocate using statistical approaches to determine the fewest possible ideal characteristics that provide the same accuracy as the whole set of input data. KNN is a policy-based machine learning model that uses four traffic metrics to identify DDoS attacks in the IoT: the number of unique IP addresses at the destination and the maximum, minimum, and mean packet counts per destination [75]. IoT devices are believed to create random IP addresses when attacked and transmit malicious data. They detected attacks 94 percent of the time with their classifier. An alternative method, based on a single DNS record [37], was offered as a solution to the time-consuming feature selection process. According to the authors, the best way to identify Mirai-like botnets in IoT networks is to examine a single DNS record at the IoT transportation layer. Using a statistical approach to feature selection, another study looked at the timestamp of the request, calculated the time difference between the previous and subsequent requests, and generated statistical characteristics like mean, standard deviation, and so on [48]. According to the authors, a standard deviation greater than the mean indicates abnormal traffic.

Without explicitly picking features to understand how these data are connected, deep learning classifiers perform well on huge datasets [20]. According to some researchers, deep learning-based models have advantages over traditional models regarding feature selection. In one study, the authors chose features based on multiple objectives (NSGA-ii-aJG) [19]. They were able to identify DDoS activities with 99.03 percent accuracy by using CNN+LSTM. Using the feature selection process, the authors were able to reduce training time by fivefold after running the same model on the entire feature set. In another study [84], deep learning techniques automatically identified characteristics and classified raw data to detect irregular traffic patterns. The suggested LSTM model scored 99.98 percent accuracy, whereas the SVM model scored 88.18 percent. In deep learning, CNNs are frequently used to select features automatically. In IDS, CNN is used to extract relevant characteristics from raw network data and then applied to image processing. In image processing applications, CNN is often applied to image processing applications. Feature selection is a labor-intensive, time-consuming, difficult, and error-prone process in machine learning [84]. It is important to have a deep understanding of network traffic to discover realistic and linked aspects. An additional constraint on the execution of feature selection is its independence from the training phase. Due to this, they do not occur simultaneously, so the classifier can perform better [84]. Deep learning overcomes these limitations by automatically detecting underlying patterns and correlations within the data set for optimal performance. A feature selection algorithm enhances significant characteristics and diminishes irrelevant ones in deep learning. IoT threat categories are discovered and classified simultaneously using feature selection and classification.

## 9. Challenges and Research Trends

The following section outlines current developments in IoT security, some of the gaps identified in the reviewed research publications, and IoT security problems. We expect that it will aid researchers in developing a robust, scalable, and accurate intrusion detection system for IoT devices that is capable of detecting large-scale assaults.

### 9.1 Selecting the right IDS strategy

DDoS attacks such as Mirai, discovered in 2016, remain a serious threat. By extending the attack surface and experimenting with new payloads, 63 Mirai-like variants were found in July 2019. The detection of botnets is crucial, and the defense of vulnerable devices requires multiple layers of defense [43]. Signature-based prevention is widely used in network intrusion detection systems and anti-virus software. A knowledge base containing existing attack signatures is required for this strategy. Using this strategy is time-consuming, needs continuous database updates to reflect newly detected malware, and propagates the updates to all locations where the database is used. These studies employ a signature-based strategy, which has been shown to have weaknesses [75]. Using heuristics and behavioral approaches, the limitations of the signature-based approach may be overcome. Heuristic techniques analyze executable code to detect malware. Behavioral methods examine malware symptoms, such as requesting privileged access or trying to access restricted files.

### 9.2 Scalable IDS solutions for IoT

Using deep learning algorithms to analyze network packets is a more scalable method of detecting zero-day threats in the IoT context. Cyber-attacks today are smart, sophisticated, and developing at a breakneck pace [10]. By analyzing available data and user behavior, deep learning algorithms uncover hidden patterns that can be used to identify malicious attempts. Researchers have proposed offloading computational tasks to edge/fog nodes due to the resource constraints of IoT devices and the computational costs of deep learning models. Business workloads and high-cost computing operations will be moved to the cloud or network edges by 2021 [27]. The IoT model must be connected to edge data centers to be effective and safe [41]. It will also provide redundancy and failover advantages in the event of a large-scale cyberattack on a critical site through the firewall, complicated model training, and intrusion detection systems.

## 9.3 Selecting the Right Dataset

Model development and training for intrusion detection must consider the unique characteristics of the IoT environment. The reduced memory, processing power, and energy of IoT devices and the increased network traffic they create require solutions that differ from those used by traditional computer systems. The IoT dataset must include various attack types to train machine learning and deep learning models. Datasets for training an IDS are available today in many forms; however, not all are current or include IoT traffic. It is necessary to pay special attention to models with unbalanced classes or fewer attack labels since they will suffer from poor performance.

## 9.4 Training and Labeling Continuously

To develop an optimum deep learning and machine learning model, you must train the model on a benchmark dataset. To optimize their performance, the researchers train machine learning models on smaller datasets when large datasets. Similarly, a network traffic model that does not consider all traffic patterns might result in a biased IDS model. Although such models may initially produce superior performance metrics, they would fail to generalize previously unknown patterns in a real-world environment. It takes a lot of time to categorize such a large dataset. In this paper's section on "data analysis," we demonstrate that deep learning is an effective strategy for identifying meaningful patterns in huge datasets. It can identify meaningful patterns without human labeling. IDS models need to be regularly updated with new network data to detect threats and accurately distinguish benign from malicious traffic.

## 10. Conclusion

Every human being has access to related items. Through gimmicks, the Internet of Things alters people's lives. Internet of Things applications include smart cities (such as smart parking), smart environments (such as noise pollution), smart meters (such as smart grids), and industrial controls. All sectors are affected, even those particularly vulnerable, such as the military, healthcare, and construction. Unfortunately, firms rely on inventiveness and increasingly integrated products instead of consistency and proof of durability. The Internet of Things is a double-edged sword. Humans can hack and exploit this connected army. An infected node can affect the entire IoT network. An unsuspecting person might be hacked into to take their automobiles via remote control by a malevolent person. In the future, IoT-based research will yield more results. The validation strategy needs to be refined, namely creating a public data set as a baseline for network sharing with IoT devices. A systematic analysis of the use of deep learning methods for the security of IoT-based systems was conducted in this work.

We presented a full taxonomy of DL approaches and discussed the context of IoT security. The second part of the paper discusses the methodology we used, beginning with elaborating the queries generated sequentially from the research questions and progressing through several screening and refining procedures to arrive at the final collection of 69 studied articles. Lastly, we categorized the findings based on the study topics into three broad categories. The security issues addressed, the DLN architectures used, their application areas, and the datasets used were all included. Our closing debate shed some light on several difficulties we have yet to resolve on the subject we chose, demonstrating that more research efforts will be necessary in the near future to make the use of DL a permanent and mature solution to IoT security. We are of the opinion that this is true. Different Internet of Things protocols may be used for other attacks. Comparing the various constructed NIDS is easy, realistic, and practical using this data collection method. In addition, IoT NIDS must detect both known and unknown attacks without being protocol dependent. Finally, NIDS designs for IoT should combine edge computing and fog computing strategies extensively. It is possible to detect IoT intrusions with these techniques with little effort.

**Funding Statement**

In this paper, the authors did not receive funding from any institution or company and declared that they do not have any conflict of interest.